\documentclass[final,5p,times,twocolumn]{elsarticle}

\usepackage{dblfloatfix}
\usepackage{euscript}   
\usepackage{graphicx}
\usepackage{dcolumn}
\usepackage{bm}
\usepackage{amsmath}
\usepackage{mathptmx}
\usepackage{textcomp}
\usepackage{tabularx} 
\usepackage{times}
\usepackage{wrapfig}
\usepackage{caption}
\usepackage{subcaption}
\usepackage{adjustbox}
\usepackage{listings}
\usepackage{silence}
\usepackage{wasysym}
\usepackage{titlesec}

\usepackage{graphicx}
\usepackage[utf8]{inputenc}
\usepackage[T1]{fontenc}
\usepackage{etoolbox}
\usepackage{csquotes}
\usepackage{xcolor}

\usepackage{hyperref}
\usepackage{url}
\hypersetup{
    colorlinks=true,
    linkcolor=green,
    filecolor=magenta,      
    urlcolor=cyan}

\usepackage{diagbox}

\journal{Journal of Physics: Materials}

\begin{document}

\begin{frontmatter}

\title{Electronic Conductivity in Metal-Graphene Composites: The Role of Disordered Carbon Structures, Defects, and  Impurities}

\author[OU]{K. Nepal}
\author[OU]{C. Ugwumadu}
\author[OU]{A. Gautam}

\author[PNNL]{Keerti Kappagantula}

\author[OU]{D. A. Drabold \corref{cor}}
\ead{drabold@ohio.edu}

\address[OU]{Department of Physics and Astronomy,
Nanoscale and Quantum Phenomena Institute (NQPI)\\
Ohio University, Athens, OH 45701, USA}

\address[PNNL]{Pacific Northwest National Laboratory, 
Richland, Washington, 99352, USA}

\cortext[cor]{Corresponding author}

\begin{abstract}
This paper explores the transport properties of aluminum-carbon composite material via \textit{ab initio} methods. Interfacial and electronic dynamics of the aluminum-graphene interface structure were investigated using models of amorphous graphene added to an aluminum matrix. We examine the impact on electronic conduction caused by the presence of nitrogen impurities within the interfacial amorphous graphene layer. We elucidate the conduction mechanisms by using a projection of the conductivity into space.
\end{abstract}


\end{frontmatter}


\section{\label{sec:introduction}Introduction}

Recent experimental studies have shown that the addition of single- or multi-layered graphene sheet(s) to metals like aluminum (Al) and copper (Cu) produces unconventional electrical wires with enhanced electronic conduction \cite{MSE_2023, Keerti_2022, DAD_175, Tokutomi, Chyada}. The increased conductivity observed in these new metal/carbon composites defies traditional principles used to understand conductivity in materials, as exemplified by Matthiessen's rule. While experimental evidence that supports this phenomenon exists, a physical explanation for the mechanism is still needed. A first attempt to explain the improved conductivity observed in copper- and aluminum-graphene composites, indicated that the enhanced conductivity can be attributed to a guiding principle: the establishment of an interfacial coverage by graphene sheet(s) serving as a conduit for electron transport across metal grains (see detailed in References \cite{apl_2023} and \cite{AM_2023}). \\

The preceding conclusions lead to two pertinent questions: (1) How does this phenomenon manifest itself in metals other than copper and aluminum? (2) To what extent must the layered carbon material's purity and crystallinity be maintained to attain these improvements? The second question is addressed in this paper. The question highlights the practical challenges associated with large-scale graphene production. The process of graphite exfoliation frequently results in defective graphene, while the synthesis of high-quality graphene is often limited to small quantities through methods such as chemical vapor deposition \cite{lin2019synthesis, hernandez2008high}. On the other hand, ultra-fast Joule heating offers the potential for generating more substantial quantities \cite{JH,JHreview}, but the method still grapples with cost and high energy consumption issues. Even the "crystalline" CVD graphene procured commercially is never completely crystalline. At best, it has crystalline domains with amorphous sections that are considered grain boundaries. Therefore, to completely describe metal/CVD-graphene behavior, we model the metal/amorphous-graphene interface.\\

In this work, the potential of amorphous graphene (aGr) to ameliorate scattering at grain boundaries is studied. Amorphous graphene is a topologically disordered carbon structure that exhibits distinct and intriguing structural, mechanical, and electronic characteristics, perhaps enabling innovative material design and technological advancements \cite{DAD_AGraphene,aGr2,aGr3, aGr4,nature_2023, prl_raj,GlassyC1}. Our proposed mechanisms reveal reduced scattering at grain boundaries dressed by pure graphene exhibiting how aGr integrates effectively into the interface. Additionally, the implications of incorporating nitrogen into both crystalline and amorphous graphene within an interface were examined.\\

The rest of this paper is organized as follows: Section \ref{sec:theory} provides a summary of the theory behind space projected conductivity technique used for the electronic conductivity calculations \cite{subedi_pssb}, while the methods used to construct the aluminum/layered carbon composites is discussed in Section \ref{sec:methods}. In Section \ref{sec:results}, the interface signatures and electronic transport in aluminum/amorphous graphene composite are analyzed in sub-section \ref{sec:structure}, and the aluminum/nitrogen incorporated amorphous graphene composite is discussed in sub-section \ref{sec:impure}. 

\section{\label{sec:theory}Theory}


The statistical mechanics of linear response, as explored by Kubo \cite{KGF1}, provides a framework for studying how a material responds to external perturbations. The expression for electrical conductivity within a single-particle approximation for wavefunctions is known as the Kubo-Greenwood formula (KGF)  \cite{KGF1,KGF2,KGF3}. The average of the diagonal elements of the conductivity tensor 
   ($\sigma_{\alpha\alpha}$), ($\alpha$ is a Cartesian coordinate index $(x, y, z )$), which for any frequency $ \omega $ is :
\begin{multline}\label{kgf}
    \sigma(\omega) = \frac{2\pi e^2}{3m^2\Omega \omega} {\sum_\textit{\textbf{k}} w_\textbf{k} \sum_{i,j} \sum_{\alpha} [ f(\epsilon_{i,\textit{\textbf{k}}} - f(\epsilon_{j,\textit{\textbf{k}}}]} \\
    |\langle\psi_{j,\textit{\textbf{k}}}| \textit{\textbf{P}}^{\alpha}|\psi_{i,\textit{\textbf{k}}}\rangle|^2 \delta(\epsilon_{j,\textit{\textbf{k}}} - \epsilon_{i,\textit{\textbf{k}}} - \hbar \omega)
\end{multline}

Where, \textit{e} and \textit{m} respectively are electronic charge and mass, $\Omega$ is the volume of the supercell, $w_\textit{\textbf{k}}$'s are the integration weight factors for \textit{\textbf{k}}-points, $\psi_{i,\textit{\textbf{k}}}$ are single particle Kohn-Sham states, in this case,   with associated energies $\epsilon_{i,\textit{\textbf{k}}}$ for band index $i$ and Bloch vector \textit{\textbf{k}}, $\textit{\textbf{P}}^{\alpha}$ is a momentum operator along $\alpha$ direction, and \textit{f($\epsilon_{i,\textit{\textbf{k}}}$)} denotes the Fermi-Dirac distribution weight. Within a limit $ \omega \rightarrow 0$, equation \ref{kgf}  yields the DC conductivity. This approach has proven valuable in studying the transport properties of conductors, semiconductors, and other materials within tight-binding and DFT approximations \cite{Allen1987, Martin,raj_copper_tan}, and may be employed with hybrid functional estimates of excited states and energies \cite{Caravati}.\\

Utilizing the KGF to gain a microscopic understanding of the systems and the spatial distribution of charge transport, a technique called space projected conductivity (SPC) \cite{subedi_pssb} was developed. While the in-depth development and application of this technique are described in earlier publications \cite{Prasai_2018, Subedi_2022, Subedi_2019}, this paper provides a concise overview of the concept. Based upon \ref{kgf} define:
$$
g_{i,j,\textit{\textbf{k}}} = \frac{2\pi e^2}{3m^2\Omega \omega} [ f(\epsilon_{i,\textit{\textbf{k}}} - f(\epsilon_{j,\textit{\textbf{k}}}]   \delta(\epsilon_{j,\textit{\textbf{k}}} - \epsilon_{i,\textit{\textbf{k}}} - \hbar \omega)
$$
   
\noindent Next, introduce a complex-valued function: $\xi_{ij\textit{\textbf{k}}}^{\alpha}(x) =  \psi_{j,\textit{\textbf{k}}}^*(x)\textit{\textbf{P}}^{\alpha} \psi_{i,\textit{\textbf{k}}}(x)$ on a discrete real-space grid (\textit{x}) uniformly spaced in a 3D supercell with spacing \textit{h}, equation \ref{kgf} is rewritten as 
   \begin{equation}\label{kgf3}
    \sigma \approx h^6 {\sum_\textit{\textbf{k}} w_\textit{\textbf{k}} \sum_{i,j,\alpha}} \sum_{x,x'} g_{i,j,\textit{\textbf{k}}} [\xi_{ij\textit{\textbf{k}}}^{\alpha}(x)][\xi_{ij\textit{\textbf{k}}}^{\alpha}(x')]^*
    \end{equation}
    
\noindent The conduction matrix is defined as

 \begin{equation}\label{kgf4}
    \Gamma(x,x') =  h^6 {\sum_\textit{\textbf{k}} w_\textit{\textbf{k}} \sum_{i,j,\alpha}} g_{i,j,\textit{\textbf{k}}} [\xi_{ij\textit{\textbf{k}}}^{\alpha}(x)][\xi_{ij\textit{\textbf{k}}}^{\alpha}(x')]^*
    \end{equation}
    
\noindent The matrix elements of $\Gamma$ have the dimension of conductivity and summation over all grid points recovers the KGF conductivity of the system within a limit as $h \rightarrow 0$. In practice, $\textit{\textbf{k}} \ = \  \textbf{0}$ point of the Brillion zone (reasonable for large supercells)  is often employed. $\Gamma$ has several interesting features discussed in references \cite{subedi_pssb}. Summing out $x^\prime$ gives the SPC $(\zeta)$ at real-space grid points $x$.

 \begin{equation}\label{spc}
    \zeta(x) = |\sum_{x'} \Gamma(x,x')|
    \end{equation}
The modulus is taken to ensure a real value for the scalar field $\zeta$.

 \begin{figure*}[!t]
 \centering
	\includegraphics[width=\linewidth]{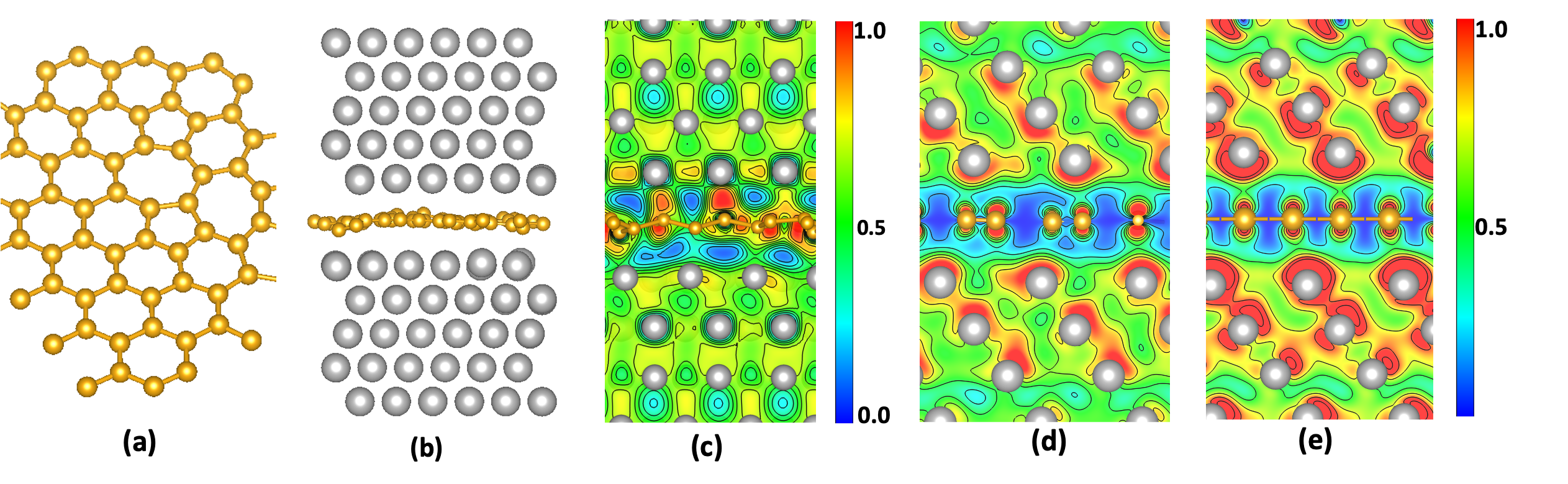}
        \caption{(a) A typical configuration of aGr with topological disorder (odd-membered carbon rings). (b) The geometry of the Al-aGr interface model. (c) 2-dimensional contour plot of electron localization function depicting mixed interfacial bonding character in the interfacial region in Al-aGr composite. Contour plot of charge density computed for twenty bands symmetrically chosen around the Fermi level for (d) Al-aGr composite and (e) Al-Gr composites along the plane transverse to the graphene. The charge density is normalized with the maximum value for the Al-Gr composite. Gold and light gray spheres are carbon and Al atoms.} 
	\label{models}
\end{figure*} 
\section{\label{sec:methods}Models and Methods}

The interface design for the aluminum-carbon composite in this work naturally evolved from earlier calculations discussed in References \cite{apl_2023} and \cite{AM_2023}. In what follows metal-graphene composite with crystalline graphene is defined as Al-Gr composites and that with amorphous graphene as Al-aGr composites. A \textit{"sandwich"} structure of Al-aGr composites was constructed by stacking a layer of aGr (see representative structure in Figure \ref{models} (a)) between two  (111) terminations of Al, as depicted in Figure \ref{models} (b). Aluminum (111) was chosen due to its low surface energy compared to other crystallographic orientations \cite{111_orient2}. The models were relaxed at constant cell volume to a configuration with minimal energy, corresponding to an average interfacial distance of $\approx$ 2.94 \r{A}. \\

First-principles calculations were conducted using the Vienna \textit{ab initio} simulation package (VASP) code \cite{K17_vasp}. The electron-ion interactions were described using the projector augmented wave (PAW) method ~\cite{K18_paw}, and the exchange-correlation functional was based on the generalized gradient approximation (GGA) of Perdew–Burke–Ernzerhof (PBE) \cite{K19_pbe}. A kinetic energy cutoff of 420 eV for relaxation and a larger cutoff of 540 eV was used for electronic properties calculations, and the system's total energy convergence accuracy was set at $1.0 \times 10^{-6} $ eV/atom. The Brillouin zone was sampled using a $2\times 2 \times 1 $ Monkhorst–Pack \textit{\textbf{k}}-mesh for optimization relaxation and for electronic structure calculations. Periodic boundary conditions were implemented throughout. The Fermi-Dirac distribution function in the KGF used a smearing temperature of 1000 K. The $\delta$ function in the expression of KGF was approximated by a Gaussian distribution function of width 0.02 eV. References \cite{parameter_1,parameter_2,parameter_3} provide a detailed discussion of various approximations inherent in the KGF formulation, such as Gaussian width, the number of atoms, smearing temperature, and energy cutoff for finite-size cells. 
\section{\label{sec:results} Results and Discussions}

\begin{figure*}[!b]
	\includegraphics[width=\linewidth]{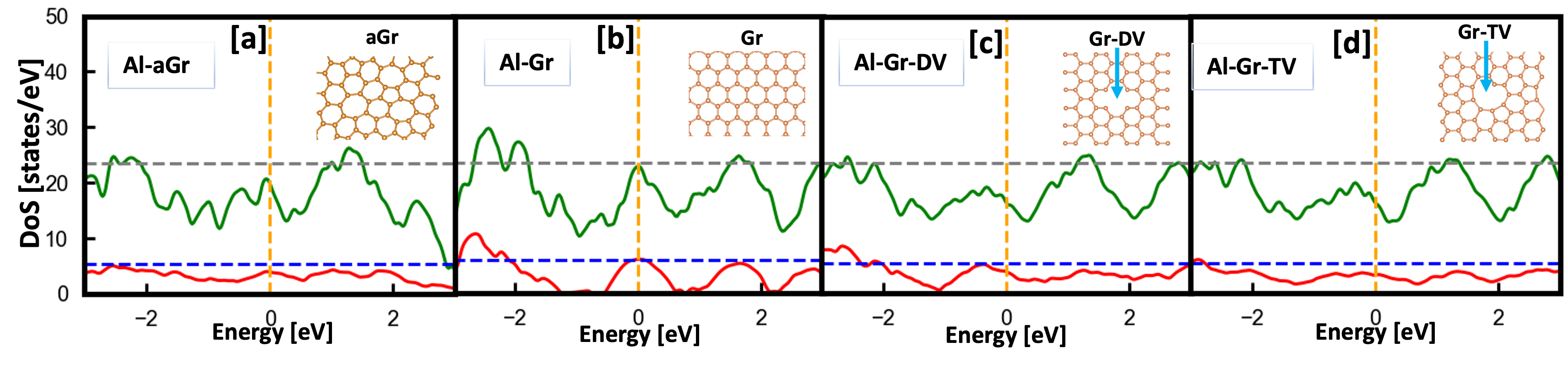}
        \caption{Total electronic density of states (green plots) and projected density of states onto carbon (C) atoms (red plots) for (a) Al-aGr composites, (b) Al-Gr composite, (c) Al-Gr composites with di-vacant graphene sheet, and (d) Al-Gr composite with a tri-vacant graphene sheet. All inset shows a representative graphene structure in each of the composites. Fermi level is shifted to E = 0, indicated by a vertical drop line. The horizontal lines, drawn in both TDoS and PDoS, serve as a guide for comparing electronic states at the Fermi level.} 
	\label{fig_2_vacant_sigma}
\end{figure*} 

\subsection{\label{sec:structure}Al-aGr Composites}

To examine the electronic dynamics at the interface region, the energy required for an electron in aluminum to surmount the potential barrier from the interface gallery to the aGr plane was calculated, yielding an energy of  $\approx $ 2.97 eV. To facilitate a qualitative comparison, a structure representing the Al-Gr interface was constructed in a similar aluminum and carbon environment. For this case, the computed energy barrier for the Al-Gr interface was $\approx$ 2.38 eV. The Al-aGr interface exhibited a higher energy barrier, likely stemming from a misaligned interface registry and the intrinsic topological disorder in the structure of the aluminum (111) surface and carbon. \\

Next, the analysis of the interfacial bonding character was carried out using the electron localization function (ELF, call it $\chi$). $\chi$ values typically span the range from 0 to 1, with 1 indicating a covalent bond, 0.5 corresponding to a metallic bond, and 0 being undefined, as defined in \cite{elf_bond}. Figure \ref{models}(c) displays contour plots of $\chi$ across (100) planes, with $\chi$ values of 1, 0.5 and 0 colored in red, yellow-green and blue, respectively. $\chi$ for the composite reveals a mixed bonding character at the interface, with both metallic and covalent bond characteristics between aluminum and carbon atoms depending on their relative positions. \\

The electronic charge distribution in the composite was analyzed from the charge density for twenty bands chosen around the Fermi level. The charge density plot along the (010) plane is shown in Figure \ref{models}(d) for the Al-aGr interface model, where the charge density increases from blue to red color. A similar calculation was done for the Al-Gr system as shown in Figure \ref{models} (e). While an interaction between the electron gas of the interfacial aluminum and the $\pi$ orbitals of sp$^2$ carbon atoms is observed for Al-aGr, it is weak compared to that of the Al-Gr system. Prior research has indicated that perfect graphene tends to establish a registry with aluminum (111) grains and enhance interfacial interaction \cite{AM_2023}. Nevertheless, observations from electron localization function and charge density indicate that incorporation of aGr aligned with aluminum grains may enhance electronic dynamics.\\

The electronic density of states for the Al-aGr composites was computed and compared to Al-Gr composites containing hexagonal graphene, as well as graphene featuring di- and tri-vacancy defects, referred to as Al-Gr-DV and Al-Gr-TV, respectively. The total density of states (TDoS; green plot) and projected density of states projected onto carbon atoms (PDoS; red plot) for the four categories are shown in Figure \ref{fig_2_vacant_sigma}. The Fermi level, indicated by the vertical drop line, has been shifted to zero. The magnitude of total DoS at the Fermi level goes as Al-Gr $>$ Al-aGr $>$ Al-Gr-DV $>$ Al-Gr-TV. \\

\begin{figure}[!t]
	\includegraphics[width=\linewidth]{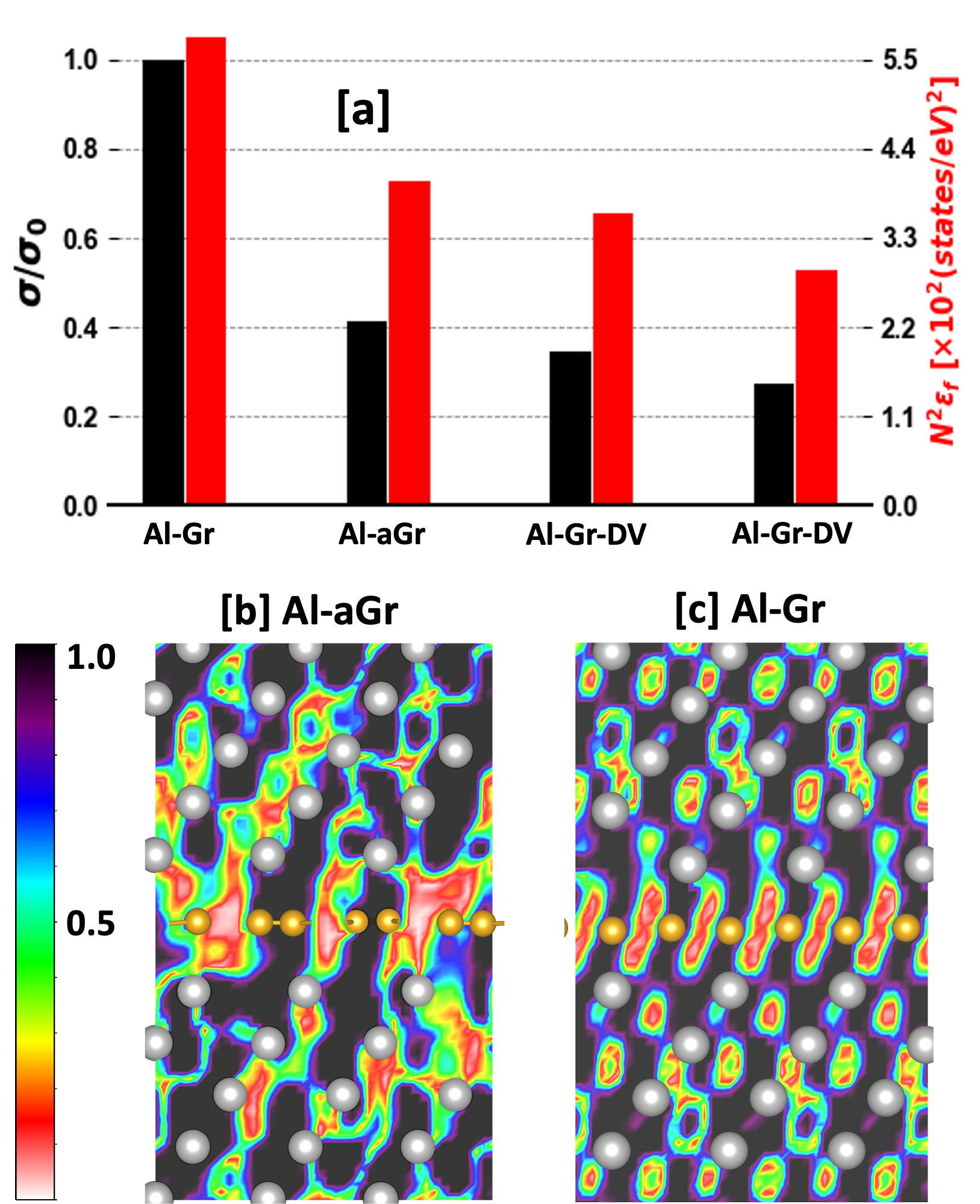}
        \caption{ (a) Average KGF Conductivities ($\sigma$) for Al-aGr and Al-Gr with perfect and defected graphene normalized by the conductivity ($\sigma_0 \approx 6 \times 10^7$ siemens/m ) of perfect Al-Gr composite, shown in left axis by black bars. The right axis with a red bar depicts the squared density of states at the Fermi level for different composites (see text). Two-dimensional normalized space-projected conductivity as heat map computed along the transverse direction (the direction normal to the plane of graphene layers) for (b) Al-aGr and (c) Al-Gr composites. The SPC is normalized with the maximum value for the Al-Gr composite. Grey (gold) spheres represent the aluminum (carbon) atoms.} 
	\label{fig_3_sigma}
\end{figure}

In line with the conventional understanding of conductivity associated with states at the Fermi level, the decrease in electronic density states correlates with a reduction in electrical conductivity ($\sigma$). As argued by Mott and Davis, the dc conductivity involves a product of the squared density of states ($N^2_{\epsilon}$) and an average of the squared momentum matrix elements over all energy states at the Fermi level ($\epsilon = \epsilon_f$), as discussed in detail in \cite{Mott_davis}. The conductivity computed for each Al-graphene interface model is shown in the bar graph presented in Figure \ref{fig_3_sigma}(a) (left axis), tracks the variation as $N^2_{\epsilon_f}$ which is shown in the right axis, depicted by red bars. For the Al-aGr composite, the average conductivity computed using KGF formalism is $\approx$ 2 × 10$^7$ Siemens/m, which is $\approx$ 40\% of the average electrical conductivity of ideal graphene Al-Gr composites. This decrease in conductivity is expected due to the reduced transport response through the topologically disordered graphene \cite{prl_raj}. However, the electrical conductivity of the Al-aGr composite is approximately 17\% higher than Al-Gr-DV and 34\% higher than Al-Gr-TV composites, respectively. This suggests that while deviation from a hexagonal graphene structure results in reduced electrical conductivity, the topological disorder is less detrimental than vacancy defect in a layered carbon structure providing insights to aGr as an alternative to realistic graphene metal composites that possess topological defects. This study is of course not exhaustive, owing to variation of defects in real interface. \\

To obtain a visual understanding of conductivity in Al-aGr, transverse SPC ($\zeta$) was computed. A two-dimensional isosurface plot for a plane along (010) is shown in Figure \ref{fig_3_sigma} (b). The $\zeta$ is normalized such that the dark region depicts high $\zeta$ values or highly favorable conduction pathways, while the reddish-light region depicts low $\zeta$ regions.  The formation of a mixed bonding environment at the interface results in graphene forming a weaker connecting link between the two Al grains on opposite sides. To make a qualitative comparison, Figure \ref{fig_3_sigma} (c) corresponding to Al-Gr composites is presented, depicting continuous conduction across graphene. In line with charge density analysis, the lack of structural alignment in Al-aGr hinders efficient transport, only facilitating partially as depicted in Figure \ref{fig_3_sigma} (b) by mixed colored regions at the interface.

\subsection{\label{sec:impure}Al-aGr with Impurities.}

Studies incorporating nitrogen into hexagonal graphene have produced material with unique electronic properties for diverse applications \cite{graphene1,graphene2,graphene3}. With the emergence of amorphous carbon structures like amorphous graphene exhibiting more intricate electronic properties, a detailed investigation into the potential of nitrogen-added amorphous carbon material holds significant interest \cite{N_doped_1, N_doped_2, N_doped_3}. Recent work to investigate the chemical process of graphitization involving coal-like composition (carbon and nitrogen, oxygen, sulfur impurities), showed that nitrogen forms a planar sp$^2$ ring substitutionally with carbon atoms at low nitrogen concentration (carbon layer formed is amorphous) \cite{Nonso_Nanotech}. This makes the investigation of materials with nitrogen-added amorphous graphene timely as the world is pushing toward an alternative to graphitic material for application. In this section, the effect of nitrogen impurities on the transport properties of Al-aGr composites is studied by substituting nitrogen into the carbon atom in the aGr layer. The model is then optimized to low energy configuration. Subsequently, the transport properties of the resulting composite system are examined.\\

  Two impurity models with $\approx$ 5\% and $\approx$ 10\%  nitrogen by weight in aGr were simulated. To make a qualitative analysis of the conductivity of these systems, Al-Gr composites with the same impurity concentrations in graphene were also studied. For both impurity concentrations, the electronic density of states was analyzed, and hence the conduction behaviors. \\

\begin{figure}[!t]
\centering
	\includegraphics[width=\linewidth]{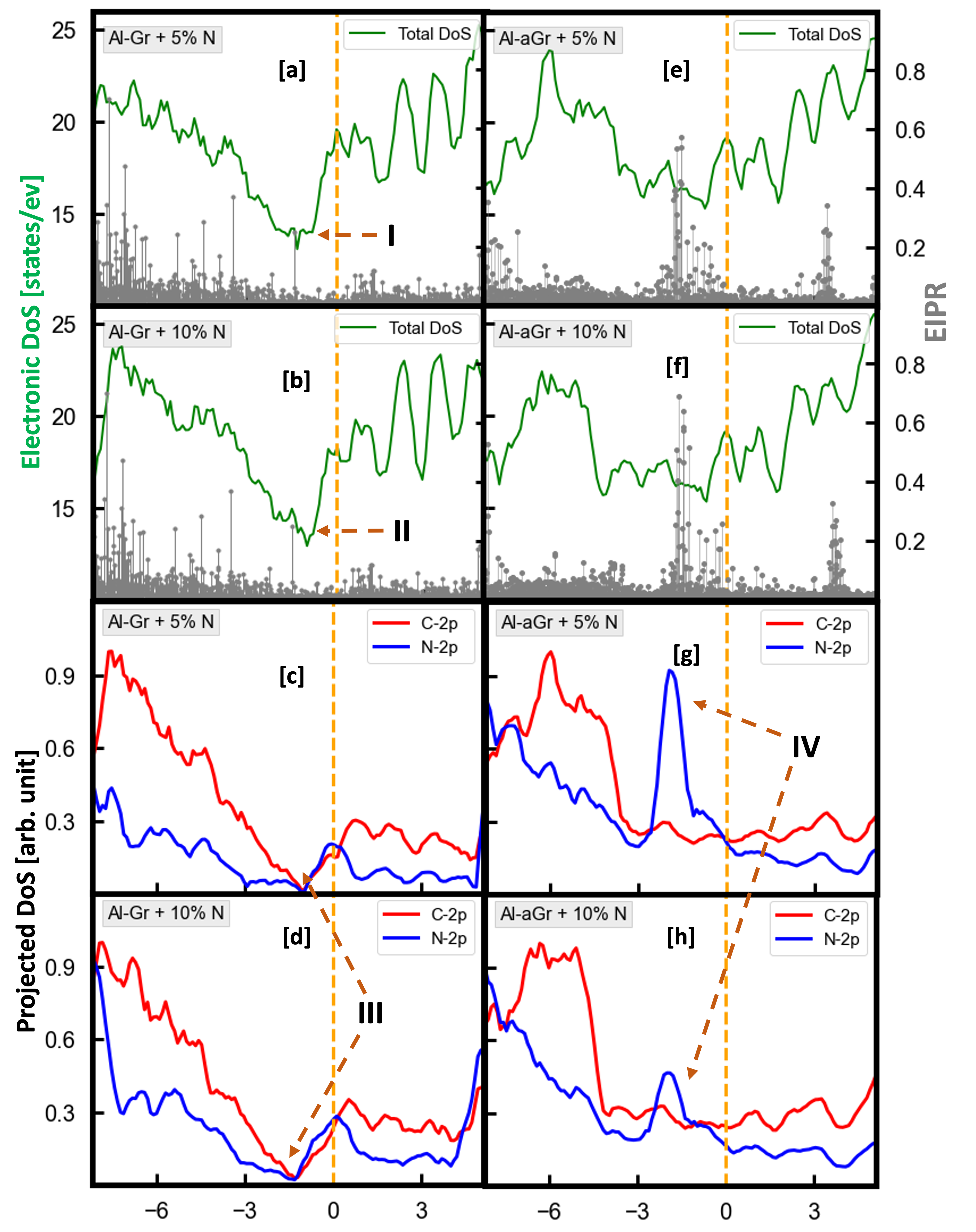}
        \caption{Total density of states (TDoS) are shown as green plots and the projected density of states (PDoS) onto valence 2p of nitrogen (N) and carbon (C) are shown by red and blue plots respectively. TDoS for Al-Gr system with (a) $\approx$ 5 \%  (b) $\approx$ 10\% of nitrogen impurity in graphene. (c) and (d) shows PDoS  for  $\approx$ 5 \% and $\approx$ 10\% nitrogen for Al-Gr systems. TDoS for Al-aGr system with (e) $\approx$ 5 \%  (f) $\approx$ 10\% nitrogen by weight in amorphous graphene. (g) and (h) represents PDoS for  $\approx$ 5 \% and $\approx$ 10\% nitrogen in Al-aGr systems. The Fermi level is set at zero, shown by vertical drop lines. Energies of particular interest [I-IV] are discussed in the text. The Electronic Inverse Participation Ratio (EIPR) for impurity models is shown on the right axis in Figures (a), (b), (e), and (f) shown by grey droplines.} 
	\label{fig_1_DOS_Vacancy}
\end{figure}

 The total electronic density of states for composites with different impurity concentrations is shown in Figure \ref{fig_1_DOS_Vacancy}. The green plots represent the total density of states, while the red and blue plots signify the density of states projected into the valence 2p orbital of carbon (C) and nitrogen (N) atoms respectively. Subplots (a) and (b) respectively depict the TDoS for $\approx$ 5 \% and $\approx$ 10\% nitrogen impurity in the Al-Gr composites while subplots (e) and (f) respectively depict for Al-aGr models. In hexagonal carbon rings, the addition of nitrogen introduces excess electrons \cite{N_doped_C1, N_doped_C2}. Consequently, the Fermi level is shifted up, shown in the left column in Figure \ref{fig_1_DOS_Vacancy}. Notice a minimum in TDoS in subplots \ref{fig_1_DOS_Vacancy} (a) and (b) labeled by I and II as an effect of the semi-conducting behavior of graphene. Carbon sp$^2$ planar rings are characterized by a prominent DoS minimum which is depicted in Figures \ref{fig_1_DOS_Vacancy} (c) and (d) labeled  III by projecting the DoS onto valence 2p orbital of carbon atoms. In contrast, a distinguishing feature in TDoS of Al-aGr composites compared to their crystalline counterpart is the filling of the minimum in DoS that was observed in Al-Gr just below the Fermi level. This is shown in Figures \ref{fig_1_DOS_Vacancy} (e) and (f). The observed electronic effect is due to the effect of odd rings in amorphous graphene leading to an increase in the density of states in the vicinity of the Fermi level, shown in Figures \ref{fig_1_DOS_Vacancy} (g) and (h). \\
 
Also, PDoS onto 2p orbital of nitrogen atoms is shown by blue plots in Figure \ref{fig_1_DOS_Vacancy}. For the case of Al-Gr, PDoS onto nitrogen shows a distinct peak at the Fermi level in par with the peak in PDoS onto carbon atoms, shown in Figures \ref{fig_1_DOS_Vacancy}(c) and (d). For nitrogen in aGr within the composite,  PDoS is characterized by an unusual peak in the energy range of -2.5 eV to -1.5 eV, labeled IV in Figures \ref{fig_1_DOS_Vacancy} (e) and (h). These observations potentially signify unique interactions or electronic configurations within Al-aGr composite material with trace nitrogen concentrations.\\

To determine the electrical consequence of nitrogen in composites, average electronic conductivities were estimated. We observe that with $\approx$ 5\% and $\approx$ 10\% nitrogen impurity in aGr, the conductivity of Al-aGr was reduced by $\approx $ 7\% and $\approx $ 12 \% respectively as compared to impurity-free Al-aGr composites.  Next with the same impurity concentrations for Al-Gr, the reduction in conductivities was $\approx $ 7.5\% and $\approx $ 20 \% compared to Al-Gr composites. The reduction in conductivities with increasing nitrogen additives is due to the introduction of localized states near the Fermi level capable of capturing and trapping charge carriers (electrons) as they move through the material, hindering their mobility and reducing the overall conductivity \cite{N_Trap}. The effect was more pronounced for Al-aGr than Al-Gr composites as observed in Figure \ref{fig_1_DOS_Vacancy} (e) and (f) (right axis), characterized by high electronic inverse participation ratio (EIPR) near the Fermi level. The EIPR measures the extent of localization of Kohn-Sham states, as discussed in references \cite{nanoporousC} and \cite{CNT}. High (low) EIPR values indicate a localized (extended) Kohn-Sham state. Further insight for such reduction from SPC analysis indicated that nitrogen atoms disrupt the electronic transport in the layered structure \cite{Nonso_Nanotech}.

\section{Conclusions}

In summary, aluminum/amorphous graphene (Al-aGr) composites and their interfacial transport properties were reported through \textit{ab initio} analysis. Building upon the enhanced electronic properties observed in aluminum-graphene composites, the potential of incorporating disordered carbon structures within aluminum grains was investigated. This work revealed mixed bonding character at the metal-graphene interface and demonstrated that aGr enables electrical conduction partially as evidenced by spatial projections of electrical conductivity. Next, with the addition of nitrogen into aGr, the average electrical conductivity of aluminum-graphene composites is hindered. These findings provide valuable knowledge for the potential of aGr as an important alternate material to ameliorate scattering at the metal grains. For completeness, we note that the present paper explores only a few configurations-specific aluminum surfaces and specific models of aG, as such the results should be viewed as qualitative. Nevertheless, this paper points the way to a more complete understanding of the remarkable behavior of metal-carbon composites.

\section*{Acknowledgments}

We acknowledge the Extreme Science and Engineering
Discovery Environment (XSEDE), supported by NSF via Grant No.
ACI-1548562 at the Pittsburgh Supercomputing Center, for
providing computational resources under Allocation No. DMR190008P, the Department of Energy Vehicles Technology Office under the Powertrain Materials Core Program, and the  Pacific Northwest National Laboratory operated by
the Battelle Memorial Institute for the U.S. Department of Energy
under Contract No. DE-AC06-76LO1830.
.


\bibliographystyle{elsarticle-num} 
\bibliography{Al_aGr}

\end{document}